# Control of valley optical conductivity and topological phases in buckled hexagonal lattice by orientation of in-plane magnetic field

Phusit Nualpijit[a] , Bumned Soodchomshom[b,*]

[a,b] Department of physics, Faculty of science, Kasetsart University, Bangkok,10900

[a] School of Integrated Science (SIS), Kasetsart University, Bangkok,10900

*Corresponding Author's E-mail: fscibns@ku.ac.th; bumned@hotmail.com

## Abstract

We investigate the optical conductivity, along with longitudinal and transverse conductivities, in buckled hexagonal lattice such as silicene subjected to both an in-plane magnetic field and a perpendicular electric field. In this model, we neglect the effect of the spin-orbit interaction, which is of a smaller order compared to the strong staggered potential and the next-nearest hoping energy. The orientations of the in-plane magnetic field and the perpendicular electric field give rise to a non-uniform, tunable gap. The Chern number for each valley degree of freedom deviates from being constant but remains steady when summed over the entire Brillouin zone. The longitudinal and transverse currents, in the case of a specific valley, can be selected by adjusting the direction of the electric field in the semimetal phase. Furthermore, the defining characteristics of topological phases induces the rapid change in longitudinal conductivity when varying the angle of orientation of the in-plane magnetic field under monochromatic light, and perfect valley filtering in transverse conductivity. The transverse current associated with a specific valley can be selected when the angle of orientation satisfies the specific conditions. This investigation paves the way for materials design with valley-locked current, using a specific orientation of the in-plane magnetic field.

## 1. Introduction

The prototype of the two-dimensional hexagonal lattice, graphene, which features two sublattices located on the same plane, has been extensively investigated due to its fascinating electronic properties [1-5]. In pristine graphene, which exhibits a zero band gap, the behavior of electrons at low energy limits has been found to follow a linear dispersion relation, akin to that of massless Dirac particles, with a Fermi velocity of $v_F \approx 10^{-6}$ m/s [1, 6]. The energy band gap can be opened due to various factors, such as strain [7, 8], and interaction with the substrate [9]. The fabrication of a high-quality, defect-free, large area has been suggested in Ref. [10]. The buckled honeycomb lattice, in which atoms within the unit cell lie on different planes, allows us to investigate the electronic properties under a perpendicular electric field [11-13]. The inclusion of the effects of a strong spin-orbit interaction (SOI) in heavy atoms such as those in silicene, germanene, tinene, and plumbene [12, 14], allows the spin-valley-dependent mass to be adjusted by a perpendicular electric field EZ. This process follows a path where the band gap closes and reopens [11] while the weak SOI in pristine graphene is negligible[15]. A magnetic substrate can induce an energy gap due to the proximity effect, as seen in silicene and germanene on WSe2. This leads to unusual performances in Xene-based spintronics devices [16-19]. The phase transition that requires the SOI can be transformed into a quantum spin Hall phase (QSH), band insulator, and quantum valley Hall phase (QVH) by varying the electric field [14, 20]. Recent reports have shown improvements in valley filtering performance related to 2D-Xene materials [21]. Moreover, the proximity effect of a magnetic substrate has been predicted to induce the quantum anomalous Hall effect (QAH)[16, 22, 23]. At the critical electric field, the spin-valley-dependent energy gaps close for two electron species while opening for the other two. The semi-metallic phase (SM) has been investigated both with and without a magnetic substrate [11, 24, 25]. Changes in phases induced by an electric field may be utilized to enhance the efficiency of transistors, enabling them to operate at lower power[26].

Haldane predicted the model of the quantum Hall effect without Landau levels [27]. Periodic magnetic flux leads to the disappearance of the total flux passing through the unit cell, a phenomenon related to the Peierls phase in the next-nearest-neighbor hopping parameter[28, 29]. Quantized Hall conductivity occurs when time-reversal symmetry is broken, resulting in non-zero dc-Hall conductivity with opposite mass signs in each valley [27]. The effects of the Peierls phase on the electronic properties of the Kagome lattice and hexagonal lattice in relation to spin orientation have been investigated[30-32]. Newly accomplished studies show the Quantum Anomalous Hall effect (QAH) with a tunable Chern number due to the orientation of magnetization in $NiAsO_3$ and $PdSbO_3$ [33]. Theoretically, the Chern number $C = \pm 1$ emerges when the magnetization lies in the x-y plane, while $C = \pm 3$ emerges when the component of magnetization is non-zero in the z-direction. This agrees with the phase in the next-nearest-neighbor hopping parameter, which allows for the calculation of non-zero Hall conductivity and Faraday rotation [32, 34].

In this paper, our model is based on a buckled honeycomb lattice that includes an in-plane magnetic field. As in the Aharonov-Bohm effect [35], the vector potential associated with the in-plane magnetic field induces a phase factor in the next-nearest-neighbor hopping parameter. The signs of mass for each nodal point can be changed by the strength and orientation of the magnetic field [36]. There is no net magnetic flux passing through the unit cell, which prohibits the emergence of a Landau level. In the absence of an in-plane magnetic field, the gap at the Dirac point remains protected. In fact, the electronic properties of a buckled honeycomb lattice can be conveniently tuned by a perpendicular electric field, which produces a staggered potential between the two sublattices. This paper focuses on longitudinal and transverse conductivities under these conditions. Furthermore, topological phases identified by the Chern number open the path for adaptation in opto-electronic devices, by utilizing a variety of phases for other 2D topological materials with the same structure. This could lead to the overcoming of new states and improvement in low-energy consumption in electronic devices [37].

## 2. Model and Methodology

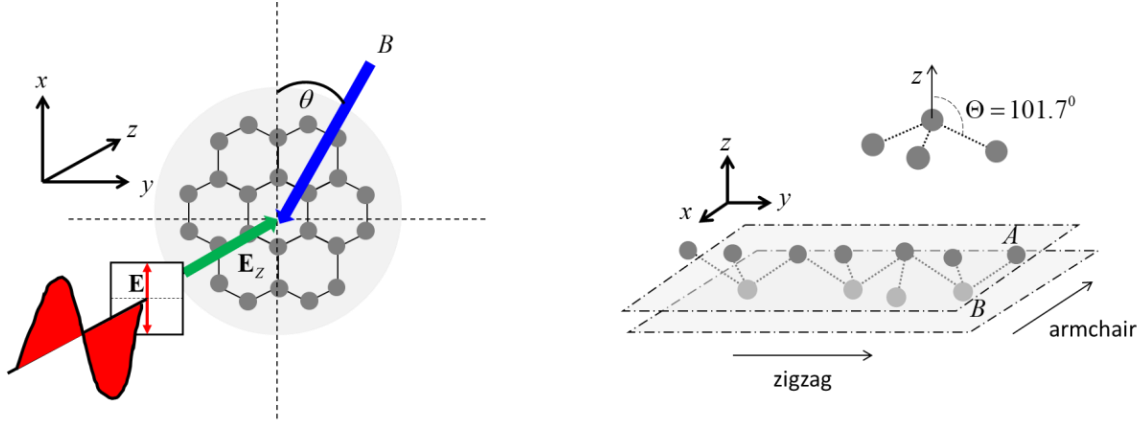

**Fig**.1 (a) The illustration of polarized light propagating in z-direction interacts with electron in the plane at $z = 0$. In this model, the in-plane magnetic field $B$ and the perperdicular electric field $E_Z$ is applied into buckled honeycomb lattice (silicene). The phase factor in the next-nearest neighbor hopping parameter has been induced due to the non-zero vector potential. Thus, the system is tunable band gap due to the perperdicular electric field and the orientation of in-plane magnetic field. (b) The side view of silicene which the angle between silicon atom and z direction is indicated by $\Theta$ which is $\Theta = 101.7^o$ [15].

### A. Hamiltonian

In this model, the two sublattices, A and B, are located in different planes due to the mixture of sp$^2$ and sp$^3$, as shown in Fig. 1b. This allows us to tune the staggered potential in the buckled honeycomb lattice using a perpendicular electric field E$_Z$, a process which is impossible in flat graphene. The standard hexagonal lattice vectors for nearest neighbor atoms, denoted as $\mathbf{a}_i$, and next-nearest neighbor atoms, denoted as $\mathbf{c}_i$, are respectively given as [38, 39]

$$\mathbf{a}_1 = a(0, -1) \qquad \mathbf{a}_2 = a/2(\sqrt{3}, 1) \qquad \mathbf{a}_3 = a/2(-\sqrt{3}, 1), \qquad (1)$$

$$\mathbf{c}_1 = \sqrt{3}a(1,0) \qquad \mathbf{c}_2 = \sqrt{3}a/2(-1,\sqrt{3}) \qquad \mathbf{c}_3 = \sqrt{3}a/2(-1,-\sqrt{3}) \, , \qquad (2)$$

where $a = 2.232$ A$^\circ$ is the bond length for silicene. Based on the Haldane model, the total magnetic flux passing through the unit cell should vanish to prevent the formation of a Landau level. This allows us to apply a uniform in-plane magnetic field, $\mathbf{B}$, for this purpose [36]. The orientation of the magnetic field with respect to the x-direction is indicated by the angle $\theta$, as shown in Fig.1a. In this case, the Landau gauge $\mathbf{A} = Bz(\sin\theta, -\cos\theta, 0)$ has been chosen, which satisfies the relation $\mathbf{B} = \nabla \times \mathbf{A}$. The second quantization Hamiltonian of the weak spin-orbit interaction in a buckled honeycomb lattice, including both a perpendicular electric field and an in-plane magnetic field, may be expressed as of the form

$$H = -t\sum_{\langle i,j\rangle\alpha} c_{i\alpha}^{\dagger}c_{j\alpha} + i\frac{\Delta_{SO}}{3\sqrt{3}}\sum_{\langle\langle i,j\rangle\rangle\alpha,\beta} \lambda_{ij}c_{i\alpha}^{\dagger}\sigma_{\alpha\beta}^{z}c_{j\beta} + \sum_{\langle\langle i,j\rangle\rangle\alpha} t'_{ij}c_{i\alpha}^{\dagger}c_{j\alpha} + \Delta_{z}\sum_{i\alpha}\xi_{i}c_{i\alpha}^{\dagger}c_{i\alpha} \, , \qquad (3)$$

where $c_{i\alpha(\beta)}^{\dagger}(c_{i\alpha(\beta)})$ is a creation(annihilation) operator of electron with spin polarized $\alpha(\beta)$ at site i taken over all pair of the nearest neighbor atoms $\langle i,j \rangle$ and next-nearest neighbor atoms $\langle\langle i,j \rangle\rangle$ with hopping parameter $t$ and $t'_{ij}$, respectively. $\sigma_{\alpha\beta}^{z}$ is the Pauli spin matrices components acting on real spin, with z. The electric field strength $\Delta_z = eE_z D$ is related to the staggered sublattice potential in which $\xi = \pm 1$ for site A(B). The second term represents the intrinsic spin-orbit interaction with strength $\Delta_{SO}$, with $\lambda_{ij} = 1(-1)$ for anticlockwise (clockwise) respect to the z-axis. In this model, we neglect the effect of the weak spin-orbit interaction (the second term may be canceled.) for instant in silicene $\Delta_{SOI} \cong 1.6 meV$, which is of a smaller order compared to the strong staggered potential and the next-nearest hopping energy $\Delta_z \gg \lambda_{SOI}$ and $t'_{ij} \gg \lambda_{SOI}$ [38]. Owing to the non-zero vector potential generated by in-plane magnetic field, the next-nearest neighbor hopping parameter has been modified by $t_{ij}' \rightarrow t'e^{\pm i\phi}$. The Peierls phase $\phi_{ij}$ which is normalized by quantum flux $\Phi_0 = h/e$ is given by the integration along the next-nearest neighbor bond from atom $i \rightarrow j$ as given by

$$\phi_{ij} = \frac{2\pi}{\Phi_0} \int_i^i \mathbf{A} \cdot d\mathbf{l} \qquad (4)$$

The integration gives rise to the total flux enclosed by two nearest neighbor bonds and one next-nearest neighbor bond. Thus, the total phase $\pm \phi_B$ should be the phases along these next-nearest neighbor bond [27, 28, 31]. In this consideration, the phase $\phi_B$ has been chosen to be the multiple of quantum flux. Using the unit vector along the vector potential $\mathbf{a}_\theta = (\sin\theta, -\cos\theta)$, and the given momentum of electron $\mathbf{k} = (k_1, k_2)$, the low energy single particle Hamiltonian in in the system may be given by

$$H = h_0 \sigma_0 + h_1 \sigma_1 + h_2 \sigma_2 + h_3 \sigma_3, \qquad (5)$$

where

$$h_0 = 2t' \sum_i \cos\left[\phi_B \mathbf{a}_\theta \cdot \mathbf{c}_i\right] \cos\left[\mathbf{k} \cdot \mathbf{c}_i\right],$$

$$h_1 = t \sum_i \cos\left[\mathbf{k} \cdot \mathbf{a}_i\right],$$

$$h_2 = t \sum_i \sin\left[\mathbf{k} \cdot \mathbf{a}_i\right],$$

$$h_3 = \Delta_z - 2t' \sum_i \sin\left[\phi_B \mathbf{a}_\theta \cdot \mathbf{c}_i\right] \sin\left[\mathbf{k} \cdot \mathbf{c}_i\right].$$

$$(6)$$

The corresponding eigenvalues and eigenvectors are respectively given as

$$E_\pm = h_0 \pm E_0 = h_0 \pm \sqrt{h_1^2 + h_2^2 + h_3^2} \qquad , \qquad (7)$$

and

$$|\psi_\pm\rangle = \frac{1}{\sqrt{2E_0(E_0 \pm h_3)}} \left(\pm E_0 + h_3, h_1 + i h_2\right) \qquad (8)$$

In this calculation of band structure shown in Fig.2, the next-nearest neighbor hopping parameter $t'$ must obey the relation $t' \ll t$ [1, 40]. The staggered potential energy due to the perpendicular electric field $\Delta_Z$, and energy of light $\omega$ are given in term of normalization by the nearest neighbor hopping parameter value $t$. The low energy band structure given in Eq.7 is shown in Fig.2. In this study, the Fermi level has been found to shift to the center between the valence band maximum and the conduction band minimum. The parameter $t' = 0.1t$ is assumed. The phase due to in-plane magnetic field $\phi_B = \pi/4$ and angle of magnetic orientation $\theta = \pi/6$ have been chosen based on a few selected values of the staggered potential energy $\Delta_Z$. This angle of orientation $\theta = n\pi/6$ gives rise to the largest energy gap by compared to other angles. This implies that the gap can open due to the staggered potential and the phase from a non-zero vector potential in this system. This is significant for control current to off-state by in-plane magnetic field in electronic junction. In the absence of an electric field, the energy gap disappears when $\theta = n\pi/3$. This happens because the electron moves along the direction of the magnetic field, leading to the inference of the absence of the Lorentz force. In this investigation, the SOI may be neglected due to the small effect comparing to the next-nearest neighbor hopping parameter $t'$ [38]. Nonetheless, the SOI should be included in strong SOI materials such as transition-metal dichalcogenides [41-43]. By varying the staggered potential energy, the critical condition occurs when

$$\Delta_Z = \sqrt{3}t'\left[\sin\left(\sqrt{3}\phi_B\sin\theta\right) + 2\cos\left(\frac{\sqrt{3}}{2}\phi_B\sin\theta\right)\cos\left(\frac{3}{2}\phi_B\cos\theta\right)\right] \equiv \Delta_c \,. \qquad (9)$$

For this condition, the gap closes at K'-point and opens at K-point. By inverting the electric field direction, the behavior of electrons around nodal points are interchanged. This condition can be used to investigate electric current flow with a specific valley.

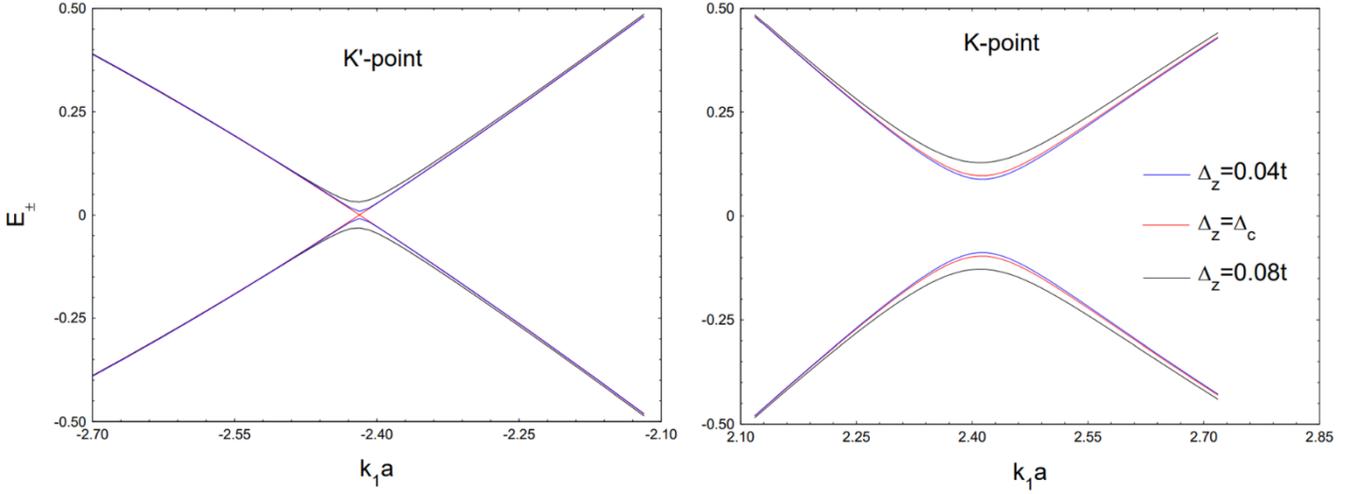

**Fig**.2 The low energy band structure around K-point and K'-points by varying the electric field strength $\Delta_z$ with fixed angle of orientation and phase from in-plane magnetic field $\theta = \pi/6$, $\phi_B = \pi/4$, respectively.

### B. Optical properties

Linear response theory is adopted to investigate the optical properties of the system. The electron is treated as a particle that interacts with an electric field from polarized light, producing current. The dynamical conductivity can be evaluated using the Kubo formula [44-46] as given by

$$\sigma_{\mu\nu}(\omega) = -\frac{i}{\hbar} \lim_{\alpha \to 0} \sum_{k \in BZ} \sum_{l,l'} \frac{\langle k,l | j_\mu | k,l' \rangle \langle k,l' | j_\nu | k,l \rangle}{E_l - E_{l'} + \omega + i\alpha} \frac{f(E_l) - f(E_{l'})}{E_l - E_{l'}}, \qquad (10)$$

where $f(E_l) = 1/\left[1 + \exp\{\beta(E_l - E_F)\}\right]$ is the Fermi-Dirac distribution function, with band index $l, l' = \pm 1$ and inverse thermal energy $\beta = 1/(k_B T)$. The current operator is given by commutation $j_\mu = -ie\left[H, r_\mu\right]$ with $\mu = x, y$. The corresponding eigenstate and eigenenergy related to Eqs.5,6 and 8 can be expressed by $|k,l\rangle$ and $E_l$, respectively. The energy of light $\omega$ is considered for ac-conductivity. The summation in k-space can be converted to integration form over the whole

Brillouin zone as $\sum_{k \in BZ} \rightarrow 1/(2\pi)^2 \int_{BZ} d^2\mathbf{k}$ [47]. The transformation $\lim_{\alpha \to 0} \alpha/\left[x^2 + \alpha^2\right] = \pi\delta(x)$ has been used for analytical calculation to separate into real part and imaginary part $\sigma_{\mu\nu}(\omega) = \sigma'_{\mu\nu} + i\sigma''_{\mu\nu}$. The matrix element of longitudinal and transverse conductivities in general form can be expressed by

$$\langle k,l | j_x | k,l' \rangle \langle k,l' | j_x | k,l \rangle = e^2 \frac{\left(E_k h_2\right)^2 + \left(h_1 h_3\right)^2}{E_k^2 \left(E_k^2 - h_3^2\right)} \tag{11}$$

$$\langle k,l | j_x | k,l' \rangle \langle k,l' | j_y | k,l \rangle = e^2 \frac{\left(h_1 + h_2\right)\left(h_3 + ilE_k\right)}{E_k^2 \left(E_k^2 - h_3^2\right)} \tag{12}$$

In low energy limit, the Hamiltonian in Eq. 5 may be approximated to

$$h_1 = k_1 \quad, \qquad h_2 = \eta k_2, \quad \text{and } h_3 = m, \tag{13}$$

where $\eta = \pm 1$ is the valley degree of freedom. The mass term $h_3$ can be created by substituting the staggered potential energy $\Delta_Z$ and by the strength and orientation of magnetic field. Thus, the analytical expression for the real part of longitudinal and transverse conductivities per spin and per valley degree of freedom, respectively given as [32, 48]

$$\sigma_{xx}^{\eta}{}'(\omega) = \frac{\pi e^2}{8h}\left[1 + \left(\frac{2|m|}{\omega}\right)^2\right]\Theta(\omega - 2|m|), \tag{14}$$

and

$$\sigma_{xy}^{\eta}{}'(\omega) = \eta\frac{\pi e^2}{8h}\left[\frac{2m}{\pi\omega}\ln\left|\frac{2|m| + \omega}{2|m| - \omega}\right|\right]. \tag{15}$$

The longitudinal conductivity disappears when $\omega < 2|m|$ relating to the Heaviside step function $\Theta(\omega - 2|m|)$ and increases rapidly at $\omega = 2|m|$. Furthermore, the function declines to the unity when $\omega > 2|m|$. The summation of spin-valley degree of freedom when $m = 0$ is nothing other than ac-conductivity of graphene in low energy limit [3]

$$\sigma_{xx}(\omega) = \frac{\pi e^2}{2h} \equiv \sigma_0 \tag{16}$$

The transverse conductivity for each valley occurs when $m \neq 0$. For $\omega \gg 2|m|$, the conductivity goes to zero, whereas for $\omega \to 0$, the function goes to the constant depending on sign of $m$ and valley degree of freedom $\eta = \pm 1$. By applying series expansion $\ln(1 + x) \approx x$ for $x \to 0$, the dc-transverse conductivity reads

$$\sigma_{xy}^{\eta}(\omega \to 0) = \text{sgn}(m)\eta \frac{e^2}{4h} \tag{17}$$

In spin independent system, the non-zero transverse conductivity occurs with the different sign of mass as discussed in Ref. [27]. Whereas in the spin-dependent system, the non-zero transverse conductivity occurs in a various condition which is related to the topological phases [24].

### C. Chern numbers

The concept of geometric phase due to the adiabatic evolution of state $|\psi(0)\rangle \to |\psi(t)\rangle$ exhibits unconventional phase which is called Berry phase [49]. This is related to Berry vector potential or Berry connection which can be defined by

$$A_\mu = -i \sum_j \langle \psi_k | \frac{\partial}{\partial k_\mu} | \psi_k \rangle \tag{18}$$

where $\mu = x, y$. The integration of gauge invarient Berry curvature $\Omega(\mathbf{k}) \equiv \nabla \times \mathbf{A}(\mathbf{k})$ over the closed surface is well-known as quantized Chern number as

$$C = \frac{1}{2\pi} \oint_C d\mathbf{k} \cdot \mathbf{A}(\mathbf{k}) = \frac{1}{2\pi} \int_A d^2\mathbf{k} \cdot \left( \nabla \times \mathbf{A}(\mathbf{k}) \right) \tag{19}$$

The relation between Chern number and dc-transverse conductivity can be expressed by [49]

$$\sigma_{xy} = \frac{e^2}{h} \frac{1}{2\pi} \int_{BZ} d^2\mathbf{k} \ \Omega(\mathbf{k}) = \frac{e^2}{h} C \tag{20}$$

This is an alternative way to identify the phases as discussed in Ref. [14, 50]. In this case the topological number has been defined by

$$C_{total} = 2(C_K + C_{K'}) \tag{21}$$

$$C_{valley} = 2(C_K - C_{K'}) \tag{22}$$

where $C_{total}$ is the total Chern number and $C_{valley}$ is valley Chern number. The factor 2 related to spin degree of freedom needs to be included to complete the reality with the spin independent model. In this system, the valley dependent Chern number for each valley may be given as

$$C_\eta = -\frac{\eta}{2} \text{sgn}(m) \tag{23}$$

where $\eta = \pm 1$ is the valley degree of freedom. When the total Chern numbers are zero, quantum valley Hall (QVH) insulator may occur. The non-trivial phases or topological phase occurs when $C_{total} \neq 0$ to get quantum anomalous Hall phase (QAH) [12, 24]. In this work we has been shown that "$h_3=m$" would depend on perpendicular electric field and strength and orientation of in-plane magnetic fields(see Eq.6).

## 3. Results and discussion

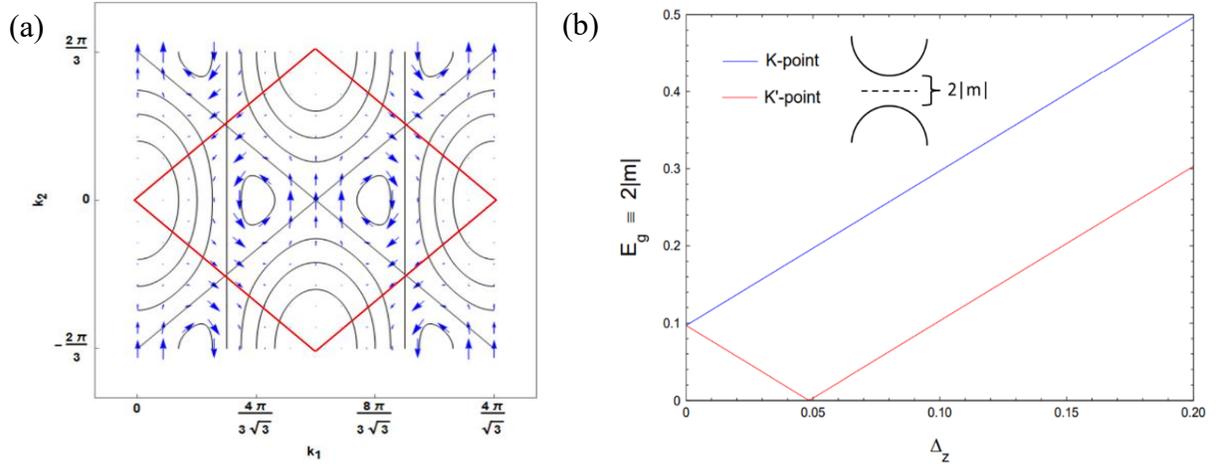

**Fig**.3  a) The area of Brillouin zone bounded by red diamond-shape and the Berry flow indicated by the blue arrows.  b) The energy gap at K-point and K'-point when varying the electric field strength $\Delta_z$ with fixed angle of the orientation and phase caused by in-plane magnetic field $\theta = \pi / 6, \ \phi_B = \pi / 4$.

We firstly show the energy gap depends on electic field in Fig.3. The perpendicular electric field and in-plane magnetic field allow us to investigate the electronic properties with tunable band gap. Furthermore, sinificantly the orientation of in-plane magnetic field induces the periodic behavior energy gap by varying angle of the orientation. Fig.3a shows the contour plot of band structure which the area of Brillouin zone is denoted by the area in the red lines. Fig.3b illustrates the energy gap by varying the electric field which the angle of orientation and phase from magnetic field have been fixed $\theta = \pi / 6, \ \phi_B = \pi / 4$. In practical terms, the calculation of the in-plane magnetic field should be very high to observe this effect. However, this prediction awaits experimental testing. In graphene, recent reports have observed strain engineering with a pseudomagnetic field [51, 52]. The straight line indicates that the energy gap consistently opens in the K-point band, while the gap can be closed and then reopened at the critical electric field in the K'-point band by varying the strength of the electric field. This effect could lead to valley-dependent transport properties.

The dc-longitudinal conductivity per spin degree of freedom in several cases of electric field are focused which shown in Fig.4. In this case, the critical electric field strength is found to be $\Delta_C = 0.0485$ which can be used to distinguish the behavior of topological phases. The quantum anomalous hall phase (QAH) satisfies the condition, given as

$$\Delta_Z < \sqrt{3}t'\left[\sin\left(\sqrt{3}\phi_B\sin\theta\right) + 2\cos\left(\frac{\sqrt{3}}{2}\phi_B\sin\theta\right)\cos\left(\frac{3}{2}\phi_B\cos\theta\right)\right].$$

The semi-metal phase (SM) satisfies the critical electric field condition, given by

$$\Delta_Z = \sqrt{3}t'\left[\sin\left(\sqrt{3}\phi_B\sin\theta\right) + 2\cos\left(\frac{\sqrt{3}}{2}\phi_B\sin\theta\right)\cos\left(\frac{3}{2}\phi_B\cos\theta\right)\right] = \Delta_c.$$

The quantum valley hall phase (QVH) occurs when the system satisfies the condition of the form

$$\Delta_Z > \sqrt{3}t'\left[\sin\left(\sqrt{3}\phi_B\sin\theta\right) + 2\cos\left(\frac{\sqrt{3}}{2}\phi_B\sin\theta\right)\cos\left(\frac{3}{2}\phi_B\cos\theta\right)\right].$$

The blue line and black line represent the longitudinal conductivity for non-zero energy gap $2m$ (recall Fig.3b) at the K-point and the K'-point $(2m_K, 2m_{K'})$ which are $(0.177t, -0.017t)$ for $\Delta_Z = 0.04t$, and $(0.257t, 0.063t)$ for $\Delta_Z = 0.08t$, respectively. The electron does not respond to the light's frequency when $\omega \leq 2|m|$ and suddenly responds when $\omega = 2|m|$ (see Fig.4). thus, the non-zero longitudinal conductivity rises rapidly at $\omega = 2|m|$ and decreases to the unit $\sigma_0$ at higher energy. On the other hand, the red line suggests that the gap perfectly closes at the K'-point and opens at the K-point $(0, 0.194t)$. The constant conductivity when $\omega \leq 2|m|$ is the same as in pristine per spin and valley degree of freedom [3, 32]. This could be adapted for **selected valley current devices without the proximity effect of magnetic strips**[53-55]. When the phase due to the in-plane magnetic field, electric field strength, and energy of light have been fixed, the plot between the conductivity and angle of orientation can be illustrated as in Fig. 5. Periodic behavior occurs, related to the tunable mass as indicated in Eq. 6. When $\Delta_z > \Delta_c$ (QVH regime), the conductivity falls and rises rapidly as shown in the black line. This agree with the result of magnetoconductivity by using Boltzmann transport theory which is the effect of chiral anomaly [56].

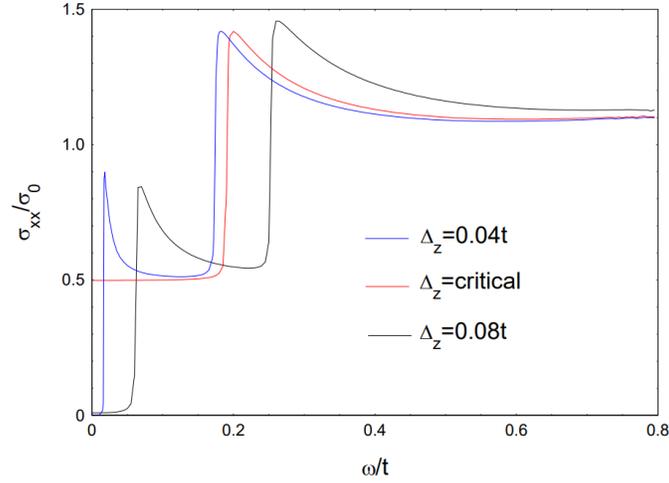

**Fig**.4   The longitudinal conductivity as a function of ligth frequency under varying the electric field strength $\Delta_z$ with fixed angle of orientation and magnetic field strength $\theta = \pi/6$, $\phi_B = \pi/4$ .

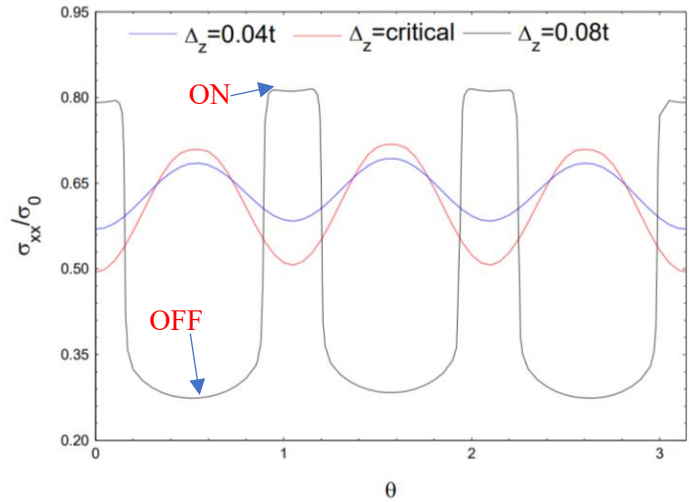

**Fig**.5   The longitudinal conductivity as a function of inplane magnetic orientation under varying the angle of orientation $\theta$ with fixed magnetic field strength $\phi_B = \pi/4$ , energy of light $\omega = 0.2$ ,

and electric field strength $\Delta_z$. Sharply switching is found for the case of $\Delta_z > \Delta_c$, the case of QSV.

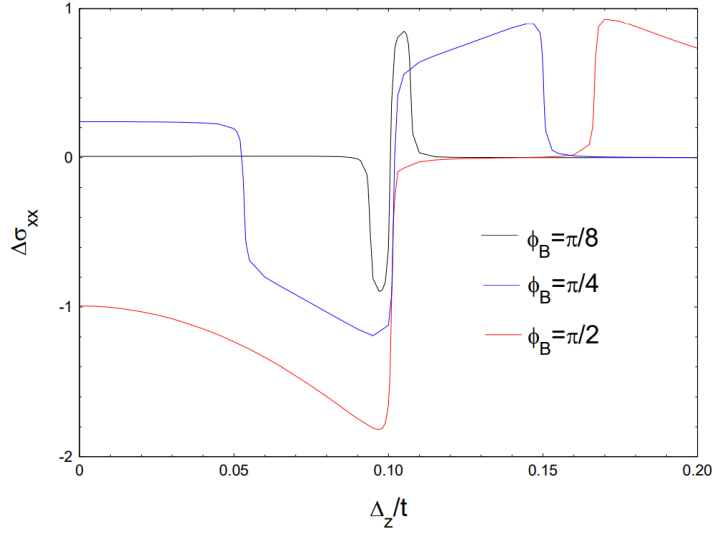

**Fig**.6 The different between maximum and minimum in longitudinal conductivity as a function of $\Delta_z$ for variouse magnetic field strength $\phi_B$ with fixed energy of light $\omega = 0.2t$.

The conductivity in Fig.5 shows a fascinating behavior which the current changes suddenly as ON (max) and OFF (min) state of a swiching transistor at some angle of orientation when $\Delta_z > \Delta_c$, which is in the QVH regime. This sharply change in current is due to magneto effect [37, 57]. In this case, the change in longitudinal conductivity would be more considered by plotting the difference between ON- conductance and OFF conductance, given by the relation

$$\Delta\sigma_{xx} = \frac{\sigma_{xx}^{\max} - \sigma_{xx}^{\min}}{\sigma_0}$$

The plot in Fig.6 illustrates the difference in longitudinal conductivity at $\theta = 0$ and $\theta = \pi/6$ with a few cases of phase $\phi_B$. The blue line indicates $\phi_B = \pi/4$, related to the result in Fig.5. The black line and red line show $\Delta\sigma_{xx}$ when decreasing and increasing the magnetic field, respectively. The in-plane magnetic-field-assisted sharply change in conductivity predicted in the model may be used as application of an optical switch [57-60]. This allows us to control the optical current by changing the orientation of in-plane magnetic field.

We next consider the transverse conductivity (or Hall conductivity) which can be given without Landau level. In this case, the Peierls phase depending on the vector potential breaking the time-reversal symmetry $\tau H(\mathbf{K}+\mathbf{k})\tau^{-1} \neq \tau H(-\mathbf{K}+\mathbf{k})\tau^{-1}$, where $\boldsymbol{\tau}$ stands for time-reversal symmetry operator in the case of QAH. The ac-transverse conductivity in cases of QAH ($\Delta_z < \Delta_c$), SM ($\Delta_z = \Delta_c$) and QVH ($\Delta_z > \Delta_c$) phases is investigated in Fig. 7. The phase $\phi_B$ and the angle of orientation $\theta$ of in-plane magnetic field have been fixed. The magnitude and direction of transverse current for each valley depends on the mass and its sign as in Eq.15 [48]. In QAH, the charge conductivity is present due to time reversal symmetry breaking. At $\omega = 0$, it is found that $\sigma_{xy}^k = \sigma_{xy}^{k'}$. In SM, the transverse current comes from single nodal point (k-valley) because of the gapless at another node, $\sigma_{xy}^k \neq 0, \sigma_{xy}^{k'} = 0$. This shows the **perfect valley filtering effect** [24, 54]. In QVH, the dc-conductivity goes to zero because of the cancelation of sign from 2 nodes, $\sigma_{xy}^k = -\sigma_{xy}^{k'}$. This behavior at $\omega = 0$, case of dc-Hall conductance, can be described based on Eq. 17. In this case, the direction of transverse current depends on the energy of light. The current disappears when the energy of light satisfies the condition

$$\left(\frac{2|m_K|+\omega}{2|m_K|-\omega}\right)^{m_K} = \left(\frac{2|m_{K'}|+\omega}{2|m_{K'}|-\omega}\right)^{m_{K'}} \, .$$

In Fig. 8a, the transverse current is found to be controllable by the orientation of in-plane magnetic field. It is always localized for $\theta = n\pi/3$ in any electric field strength and energy of light because of without Lorentz force for next-nearest neighbor jumping. Fig.8b shows the transverse conductivity for each valley. The **perfect valley filtering current** occurs only in SM phase. The **selected valley current** occurs when the angle of rotation satisfies the conditions of $\theta = \pi/6$, $\pi/2$ and $5\pi/6$. It is also found that the direction of the pure k-valley transverse current in SM phase can be controlled by varying the angle of the orientation of magnetic field (see the red curve in Fig.8b).

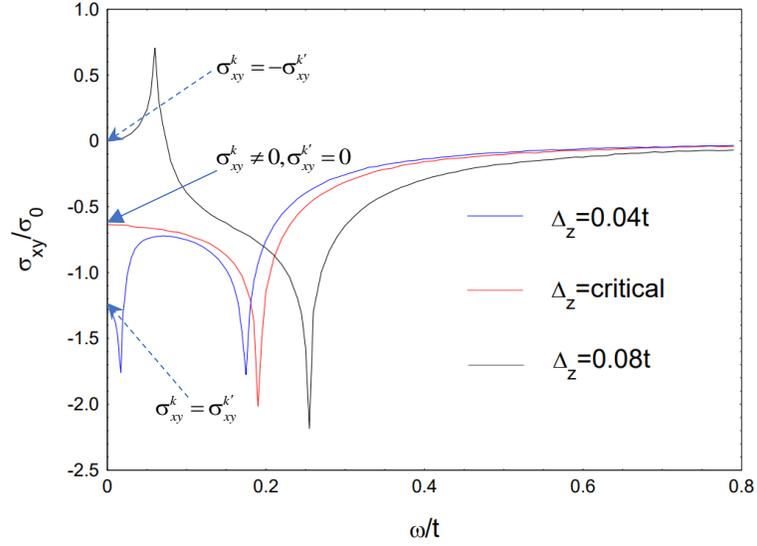

**Fig**.7  The transverse conductivity by varying the electric field strength $\Delta_z$ with fixed angle of orientation and magnetic field strength $\theta = \pi/6$, $\phi_B = \pi/4$. Complete valley filtering is found for SM phase.

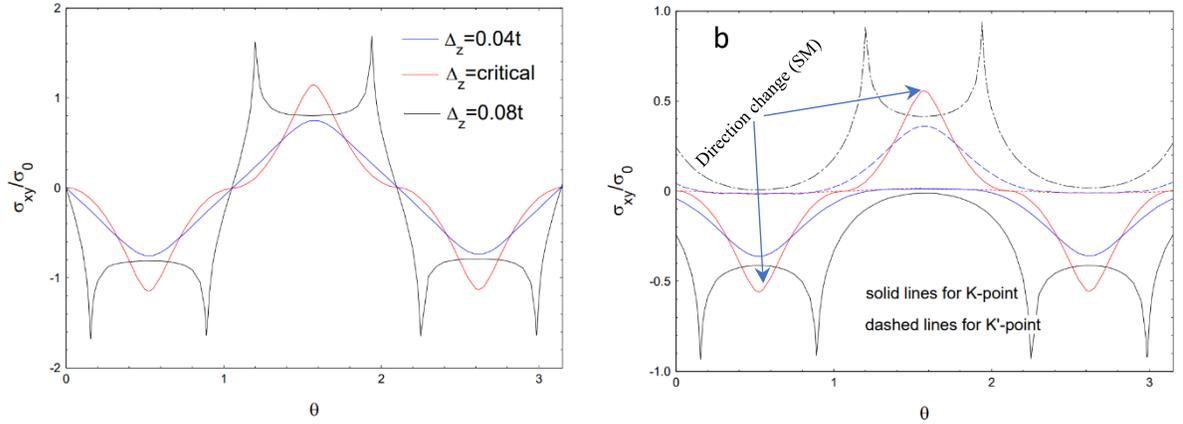

**Fig**.8  a) The transverse conductivity as a function of the angle of orientation $\theta$ at $\phi_B = \pi/4$, and $\omega = 0.2t$. b) The transverse conductivity for each valley of electric field strength $\Delta_z$ given the same as in a).

Finally, Bery curvature, Chern numbers and topological phases are investigated and discused, as seen in Figs.9, 10 and Table 1, respectively. In absent of mass term $h_3$ in Eq.6, the band structure is gapless. It has been found that the Berry curvature is very large with Berry phase $\pi$ for $h_3 \neq 0$. This can be probed by ARPES measurements [61]. The Berry flow related to Berry connection can be illustrated in the blue arrow in Fig.3a which display the different curl in different valley and becomes larger near nodal point. The plot of Berry curvature along $k_1$ direction can be illustrated in Fig.9a for K'-point and Fig.9b for K-point. Around K'-point, the negative curvature reduces and become positive curvature as increasing the electric field strength corresponding to Fig.2a. The curvature is always negative at K-point which reduces the curvature as increasing the electric field strength. This means that the band structure tend to be flat around K-point. The Chern number for each nodal point by varying electric field strength can be expressed in an inset in Fig.10. When the momentum dependent mass in $h_3$ is taken into account, the electric field strength supporting by in-plane magnetic field convinces slightly deviation from the constant for each valley but holds $C_K + C_{K'}$ to be constant. This may infer that the Chern number over the whole Brillouin zone preserves the quantized quantity but Chern number for each valley is not always true. The deviation of Chern number for each valley may occur by various reason such as in strain graphene with tunable exchange field which was discussed in [62]. The phase $\phi_B$ behave similar to the strain effect which shifts the electric field strength to change the topological phase. In the case of momentum dependent mass $h_3$ in Eq.6 being neglected, the analytical calculation of Berry curvature for the occupied band is given as of the simple form

$$\Omega(\mathbf{k}) = \frac{m}{2(m^2 + k^2)^{3/2}} \, .$$

This give rise to Chern number $C_{total} = -2$ , $C_K = C_{K'} = -1$ for $\Delta_z < \Delta_c$ and $C_{total} = 0$, $C_K = -C_{K'} = -1$ for $\Delta_z > \Delta_c$. In the critical electric field strength, the gap opens only a K-point. Thus, the Chern number $C_{total} = -1$ with $C_K = -1$ and $C_{K'} = 0$. This agree with the plot of Berry curvature which are negative except when $\Delta_z > \Delta_c$ at K'-point. The non-zero total Chern number occurs when $\Delta_z \leq \Delta_c$ in the case of fixing $\theta$ and $\phi_B$ , listed in Table 1.

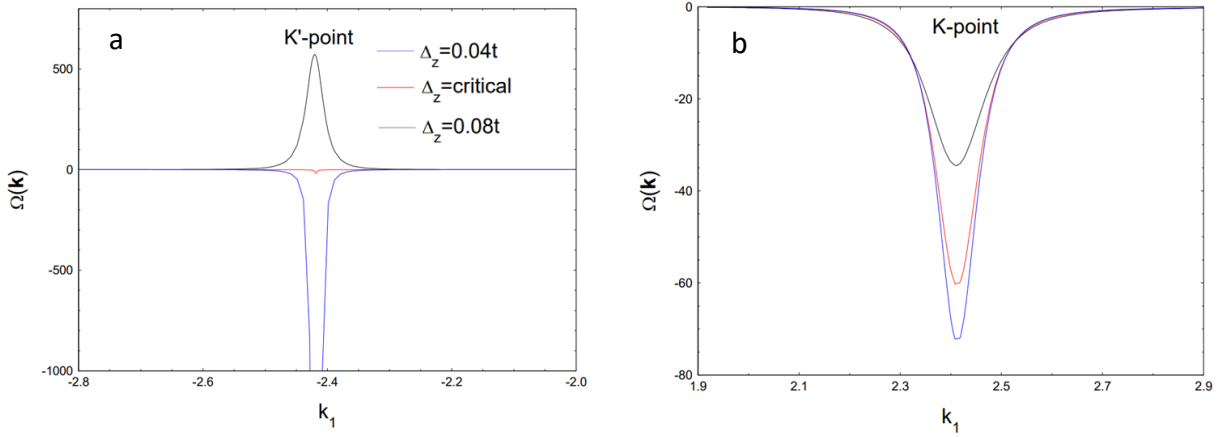

**Fig**.9   The plot of Berry curvature along $k_1$ direction around K'-point (a) and K-point (b). The given phase and orientation of magnetic field are $\theta = \pi / 6$, $\phi_B = \pi / 4$ . The blue line indicates the topological phase with non-zero Chern number. The black line is trivial phase with zero Chern number over the Brillouin zone [12].

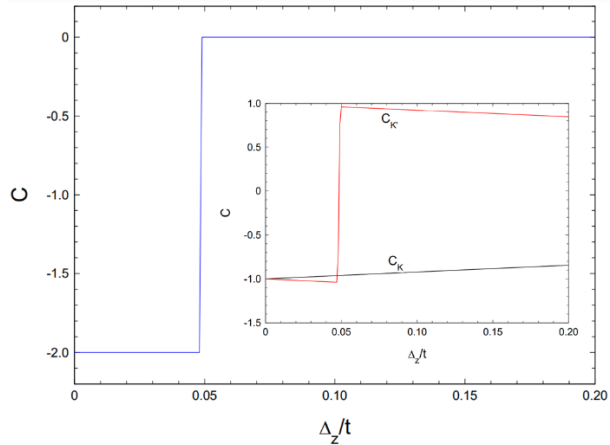

**Fig**. 10   The first (or total) Chern number C by varying the electric field strength $\Delta_z$ with fixed angle of orientation and magnetic field strength $\theta = \pi / 6$, $\phi_B = \pi / 4$ . An inset show the total Chern number for each valley $C_k$ and $C_{k'}$.

| | $C_K$ | $C_{K'}$ | $C_{total}$ | $C_{valley}$ | |
|---|---|---|---|---|---|
| $\Delta_z = 0.04t$ | −1 | −1 | −2 | 0 | Quantum anomalous Hall phase (QAH) |
| $\Delta_z = \Delta_c$ | −1 | 0 | −1 | −1 | Semimetal phase |
| $\Delta_z = 0.08t$ | −1 | 1 | 0 | −2 | Quantum valley Hall phase(QVH) |

Table 1 shows the Chern numer for each valley, total Chern number, and valley Chern number for $\theta = \pi / 6$, $\phi_B = \pi / 4$.

# 4. Summary and Conclusion

We have investigated the optical properties of in buckled hexagonal lattice such as silicene subjected to both an in-plane magnetic field and a perpendicular electric field. Furthermore, the topological phases identified by the Chern numbers induced by electric field and magnetic field has been discussed. The perfect valley filtering current has been predicted when the electric field strength equals the critical condition $\Delta_z = \Delta_c$ and $\omega \ll 2\Delta_c$. The direction of pure valley Hall current is also found to controlled by angle of orientation of the in-plane magnetic field. Moreover, when the monochromatic light interacts with electron with the condition $\Delta_z > \Delta_c$, the longitudinal conductivity changes swiftly by orientation of in-plane magnetic field, applicable for optical switch. The behavior is similar to magnetoresistance effect in silicene based controlled by magnetization configuration which is directional dependent [57, 59, 60]. The transverse conductivity without Landau level exists because the Peierls Phases breaks the time reversal symmetry. The total Chern number $C_{total} = -2$ which is the topological

phase(QAH) occurs when $\Delta_z < \Delta_c$. As the electric field strength is stronger than the critical condition $\Delta_z > \Delta_c$, the trivial insulator (QVH) with zero total Chern number appears. The perfect valley filtering effect in transverse conductivity occurs in monochromatic light with the condition of specific orientation of in-plane magnetic field $\theta = \pi/6$, $\pi/2$, $5\pi/6$. It is possible to gain advantage from these conditions which is quite constant behavior in transverse conductivity. Around these conditions, the experimental design operating for valley valve or valley locked current similar to twisted bilayer graphene may be available [63-65]. This paves the way for the materials design with spin-valley momentum locking supporting by the perfect spin-valley filtering [66], a topological quantum field effect transistor (TQFET) [26].

## Acknowledgement

This project is funded by National Research Council of Thailand (NRCT): NRCT5-RSA63002-15.